\documentclass[useAMS,usenatbib]{mn2e}

\usepackage{graphicx}
\usepackage{natbib}
\usepackage{hyperref}
\usepackage{enumerate}
\usepackage{longtable}
\usepackage{deluxetable}
\usepackage{txfonts}
\usepackage{times}

\newcommand{\hi}{{\sc H\,i}}
\newcommand{\mhi}{$M$(\hi)}

\newcommand{\mJybeam}{mJy beam$^{-1}$}
\newcommand{\msun}{{M$_\odot$}}

\newcommand{\kms}{$\,$km$\,$s$^{-1}$}
\newcommand{\ltsima} {$\; \buildrel < \over \sim \;$}
\newcommand{\gtsima} {$\; \buildrel > \over \sim \;$}
\newcommand{\lta} {\lower.5ex\hbox{\ltsima}}
\newcommand{\gta} {\lower.5ex\hbox{\gtsima}}

\newcommand{\atlas}{ATLAS$^{\rm 3D}$}

\title[A giant \hi\ tail in HCG~44]{Discovery of a giant \hi\ tail in the galaxy group HCG~44}
\author
[Paolo Serra et al.]{\parbox{\textwidth}{
Paolo Serra,$^{1}$\thanks{E-mail:\texttt{serra@astron.nl}}
B\"{a}rbel Koribalski,$^{2}$
Pierre-Alain Duc,$^{3}$
Tom Oosterloo,$^{1,4}$
Richard M. McDermid,$^{5}$
Leo Michel-Dansac,$^{6}$
Eric Emsellem,$^{6,7}$
Jean-Charles Cuillandre,$^{8}$
Katherine Alatalo,$^{9}$
Leo Blitz,$^{9}$
Maxime Bois,$^{10}$
Fr\'ed\'eric Bournaud,$^{3}$
Martin Bureau,$^{11}$
Michele Cappellari,$^{11}$
Alison F. Crocker,$^{12}$
Roger L. Davies,$^{11}$
Timothy A. Davis,$^{7}$
P. T. de Zeeuw,$^{7,13}$
Sadegh Khochfar,$^{14}$
Davor Krajnovi\'c,$^{7}$
Harald Kuntschner,$^{7}$
Pierre-Yves Lablanche,$^{6,7}$
Raffaella Morganti,$^{1,4}$
Thorsten Naab,$^{15}$
Marc Sarzi,$^{16}$
Nicholas Scott,$^{17}$
Anne-Marie Weijmans,$^{18}$\thanks{Dunlap Fellow}
and Lisa M. Young,$^{19}$}\vspace{0.4cm}\\ 
\parbox{\textwidth}{
$^{1}$Netherlands Institute for Radio Astronomy (ASTRON), Postbus 2, 7990 AA Dwingeloo, The Netherlands\\
$^{2}$Australia Telescope National Facility, CSIRO Astronomy and Space Science, PO Box 76, Epping, NSW 1710, Australia\\
$^{3}$Laboratoire AIM Paris-Saclay, CEA/IRFU/SAp -- CNRS -- Universit\'e Paris Diderot, 91191 Gif-sur-Yvette Cedex, France\\
$^{4}$Kapteyn Astronomical Institute, University of Groningen, Postbus 800, 9700 AV Groningen, The Netherlands\\
$^{5}$Gemini Observatory, Northern Operations Centre, 670 N. A`ohoku Place, Hilo, HI 96720, USA\\
$^{6}$Universit\'e Lyon 1, Observatoire de Lyon, Centre de Recherche Astrophysique de Lyon and Ecole Normale Sup\'erieure de Lyon, 9 avenue Charles Andr\'e, F-69230 Saint-Genis Laval, France\\
$^{7}$European Southern Observatory, Karl-Schwarzschild-Str. 2, 85748 Garching, Germany\\
$^{8}$Canada-France-Hawaii Telescope Corporation, 65-1238 Mamalahoa Hwy., Kamuela, HI 96743, USA\\
$^{9}$Department of Astronomy, Campbell Hall, University of California, Berkeley, CA 94720, USA\\
$^{10}$ LERMA, Observatoire de Paris and CNRS, 61 Avenue de lÕObservatoire, 75014 Paris, France\\
$^{11}$Sub-Dept. of Astrophysics, Dept. of Physics, University of Oxford, Denys Wilkinson Building, Keble Road, Oxford, OX1 3RH, UK\\
$^{12}$Department of Astronomy, University of Massachusetts, Amherst, MA 01003, USA\\
$^{13}$Sterrewacht Leiden, Leiden University, Postbus 9513, 2300 RA Leiden, the Netherlands\\
$^{14}$Max-Planck Institut f\"ur extraterrestrische Physik, PO Box 1312, D-85478 Garching, Germany\\
$^{15}$Max-Planck-Institut f\"ur Astrophysik, Karl-Schwarzschild-Str. 1, 85741 Garching, Germany\\
$^{16}$Centre for Astrophysics Research, University of Hertfordshire, Hatfield, Herts AL1 9AB, UK\\
$^{17}$Centre for Astrophysics and Supercomputing, Swinburne University of Technology, Hawthorn, Vic, 3122, Australia\\
$^{18}$Dunlap Institute for Astronomy \& Astrophysics, University of Toronto, 50 St. George Street, Toronto, ON M5S 3H4, Canada\\
$^{19}$Physics Department, New Mexico Institute of Mining and Technology, Socorro, NM 87801, USA\\
}}

\begin{document}

\date{Accepted .... Received ...; in original form ...}

\pagerange{\pageref{firstpage}--\pageref{lastpage}} \pubyear{2012}

\maketitle

\label{firstpage}


\begin{abstract}
We report the discovery of a giant \hi\ tail in the intra-group medium of HCG~44 as part of the \atlas\ survey. The tail is $\sim300$ kpc long in projection and contains $\sim5\times10^8$ \msun\ of \hi. We detect no diffuse stellar light at the location of the tail down to $\sim28.5$ mag arcsec$^{-2}$ in $g$ band. We speculate that the tail might have formed as gas was stripped from the outer regions of NGC~3187 (a member of HCG~44) by the group tidal field. In this case, a simple model indicates that about $1/3$ of the galaxy's \hi\ was stripped during a time interval of $<1$ Gyr. Alternatively, the tail may be the remnant of an interaction between HCG~44 and NGC~3162, a spiral galaxy now $\sim650$ kpc away from the group. Regardless of the precise formation mechanism, the detected \hi\ tail shows for the first time direct evidence of gas stripping in HCG~44. It also highlights that deep \hi\ observations over a large field are needed to gather a complete census of this kind of events in the local Universe.
\end{abstract}


\section{Introduction}
\label{sec:intro}

Deep observations of neutral hydrogen (\hi) around galaxies often reveal faint gaseous distributions which do not trace galaxies' stellar light. Known cases include extremely large \hi\ discs and rings around both late-type galaxies \citep[e.g.,][]{1984AJ.....89.1319K,1996AJ....111.1551M,2009MNRAS.400.1749K} and early-type galaxies \citep[e.g.,][]{1991A&A...243...71V,2006MNRAS.371..157M,2007A&A...465..787O,2012MNRAS.422.1835S}, tidal tails around merger remnants \citep[e.g.,][]{1994ApJ...423L.101S,2003MNRAS.339.1203K,2011MNRAS.417..863D}, tidal and ram-pressure tails in galaxy groups and clusters \citep[e.g.,][]{2001ASPC..240..867V,2004MNRAS.349..922D,2005A&A...437L..19O,2010AJ....139..102E,2012MNRAS.419L..19S,2012MNRAS.422.1835S}, gas accretion signatures \citep[][and references therein]{2008A&ARv..15..189S} and objects whose origin is still debated \citep[e.g.,][]{1983ApJ...273L...1S,2009Natur.457..990T,2010ApJ...717L.143M}. These systems carry invaluable information on the way galaxies assemble their stellar mass and accrete and lose gas in different environments.

Neutral hydrogen distributions revealing on-going tidal interaction, galaxy merging and gas stripping are particularly common in groups of galaxies (e.g., \citealt{1979A&A....75...97V,2001A&A...377..812V,2008MNRAS.384..305K,2009AJ....138..295F}; see also \citealt{2001ASPC..240..657H} and references therein). The observation of these processes \it in action \rm suggests that the morphology and gas content of galaxies can undergo substantial evolution inside a group. In fact, group processes might be a major driver of the morphology-density relation (\citealt{1980ApJ...236..351D,1984ApJ...281...95P}; for a discussion of the role of groups in determining this relation see, e.g., \citealt{2009ApJ...692..298W,2011MNRAS.415.1783B,2011MNRAS.416.1680C}) and of the decrease of galaxies' \hi\ content with increasing environment density \citep[e.g.,][]{2001A&A...377..812V,2009MNRAS.400.1962K,2012ApJ...747...31R,2012MNRAS.422.1835S}.

Strong indications of the importance of group processes come from a combination of optical and \hi\ results presented recently as part of the \atlas\ survey \citep{2011MNRAS.413..813C}. Firstly, the fraction of fast rotating early-type galaxies increases (and the fraction of spiral galaxies decreases) with environment density following a surprisingly tight log-linear relation, which is steeper and better defined when the density is measured on a group-like scale (defined by the distance from the third closest neighbour) rather than a cluster-like scale \citep[tenth neighbour;][]{2011MNRAS.416.1680C}. Secondly, the \hi\ morphology of gas-rich early-type galaxies appears to be strongly related to environment density when the latter is measured on a group-like scale \citep{2012MNRAS.422.1835S}. Large, regular, settled \hi\ distributions are typical in poor environments (where the distance from the third neighbour is larger than a few Mpc). More disturbed distributions are typical in galaxy groups and may be revealing the processes responsible for the morphology-density relation.

We note that understanding group processes is important to understand galaxy properties over a broad range of \it large-scale \rm environment densities. On the one hand, in a $\Lambda$CDM Universe galaxy clusters grow by accretion of groups of (rather than isolated) galaxies, and pre-processing in groups may be important to shape the properties of galaxies living in clusters at redshift zero. On the other hand, even inside large-scale voids galaxies live clustered in small groups and their evolution is to some extent driven by tidal interactions and merging \citep{1996AJ....111.2150S,2012AJ....144...16K}.

\begin{figure}
\includegraphics[width=8.4cm]{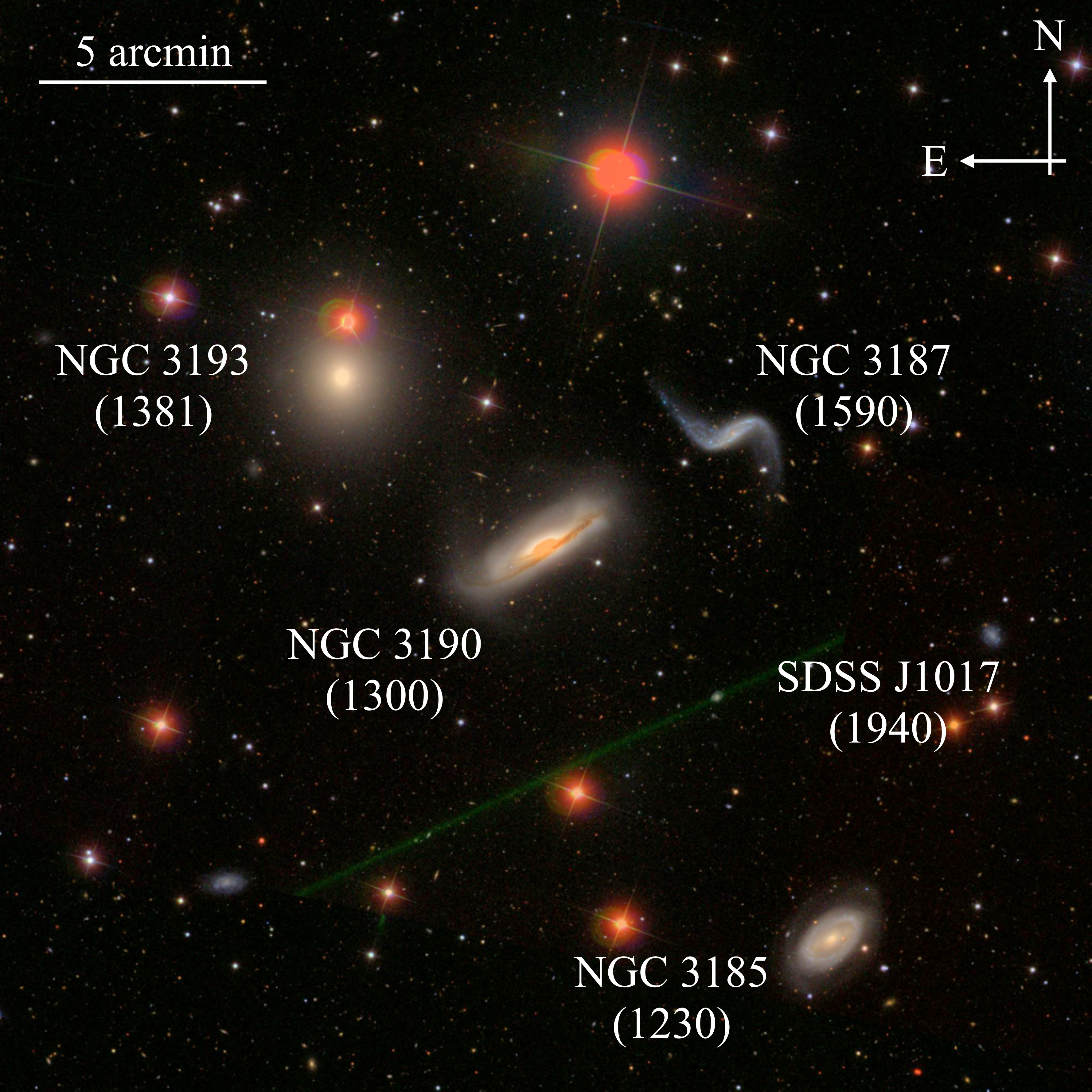}
\caption{Sloan Digital Sky Survey (SDSS) optical colour image of HCG~44 (data release 8). The image covers an area of $0.4\times0.4$ deg$^2$. We obtained the image at http://skyserver.sdss3.org/dr8/en/tools/chart/chart.asp. For each galaxy we indicate in parenthesis under the galaxy name the heliocentric recessional velocity in \kms\ (see Table \ref{tab:memb}).}
\label{fig:fig00}
\end{figure}

Within this context, stripping of gas from galaxies in groups plays a key role. Although direct detection of stripped gas has been possible in a number of groups (see references above), in many other systems galaxies are found to be \hi\ deficient but no intra-group \hi\ is detected. From an observational point of view the main challenge is that the stripped gas can diffuse quickly in the group medium (in about a group crossing time). It therefore reaches low column densities, requiring high sensitivity, and spreads over large areas, requiring observations over a large field. At the moment it is therefore unclear how ubiquitous gas stripping is in groups and, in more quantitative terms, what fraction of the original \hi\ mass of a galaxy can be stripped and on what timescale. These estimates are needed to understand the importance of gas stripping relative to other processes which may play a role in determining the morphology-density relation, such as the decrease of the cold-gas accretion rate in denser environments.

In this article we report the discovery of a giant \hi\ tail in the galaxy group HCG~44 as part of the \atlas\ survey. We summarise the properties of the group in Sec. \ref{sec:hcg44}, describe radio and optical observations in Sec. \ref{sec:obs}, present and discuss the results in Secs. \ref{sec:res} and \ref{sec:dis}, and draw conclusions in Sec. \ref{sec:con}.

\section{Hickson Compact Group 44}
\label{sec:hcg44}

HCG~44 is a compact group at a distance of $\sim25$ Mpc (see below) hosting four galaxies of comparable magnitude within an area of $\sim15\times15$ arcmin$^2$ \citep[Fig. \ref{fig:fig00};][]{1982ApJ...255..382H}. The two galaxies in the centre of the figure, NGC~3187 and NGC~3190\footnote{Some authors refer to this galaxy as NGC~3189. According to the NASA Extra-galactic Database NGC~3189 is the south-western side of NGC~3190.}, exhibit signs of tidal interaction. The former is a blue, late-type spiral with long tails pointing $\sim90$ degrees away from the plane of the galaxy. The latter is an earlier-type system with a disturbed morphology and a prominent dust lane. \cite{1959VV....C......0V} groups these galaxies together in the object VV~307 (later catalogued as Arp~316), suggesting that they are interacting. The early-type galaxy to the north-east, NGC~3193, is very close in projection to these two galaxies and has regular morphology. To the south-west, NGC~3185 is a barred galaxy with a warped, star forming outer ring. In addition to these four galaxies listed in the original \cite{1982ApJ...255..382H} catalogue, the much smaller galaxy SDSS~J101723.29+214757.9 (hereafter, SDSS~J1017) might be at larger distance (see below), while the small blue galaxy in the south-east corner is definitely a background object based on its SDSS redshift of 0.014.

The group membership is not well established. Table \ref{tab:memb} lists recessional velocity and distance estimates of the galaxies mentioned above. NGC~3185, NGC~3190 and NGC~3193 have comparable velocities. Relative to these, the velocity of NGC~3187 seems too large for such a small group. SDSS~J1017 has even higher velocity, casting doubts on its membership.

\begin{table}
{\centering
\caption{Galaxies in the HCG~44 field}
\begin{tabular}{rrrrr}
\hline
galaxy & $v_\mathrm{hel}$ & distance & method & reference\\
  & (km s$^{-1}$) & (Mpc) & \\
(1) & (2) & (3) & (4) & (5) \\
 \hline
NGC~3185     & 1230 & $20\pm5$ & TF & $a$ \\
NGC~3187     & 1590 & $26\pm10$ & TE & $b$ \\
NGC~3190     & 1300 & $24\pm5$ & SNIa & $c, d, e, f, g, h, i$ \\
NGC~3193     & 1381 & $34\pm3$ & SBF & $j$\\
SDSS~J1017 & 1940 & $29$ & HF & this work\\
\hline
\label{tab:memb}
\end{tabular}

}

\it Column 1. \rm Galaxy name. \it Column 2. \rm Heliocentric velocity measured from the \hi\ data discussed in this article except for NGC~3193, for which we use the value given by \cite{2011MNRAS.413..813C}. \it Column 3. \rm Redshift-independent distance for all galaxies except SDSS~J1017, for which we assume Hubble flow (see below). Error bars are taken from the references in column 5 except for NGC~3190, for which we give the range of distances obtained by different groups who studied the two SNIa 2002bo and 2002cv. \it Column 4. \rm Method used to determine the distance: TF = Tully-Fisher relation corrected for Malmquist bias; TE = Tully estimate; SBF = surface brightness fluctuation corrected for Malmquist bias; HF = Hubble flow with $h=0.73$ after correction for Virgo infall by $+200$ \kms. \it Column 5. \rm References for the distance estimate: (a) \cite{2009ApJS..182..474S}; (b) \cite{1988ngc..book.....T}; (c) \cite{2005ApJ...624..532R}; (d) \cite{2006ApJ...645..488W}; (e) \cite{2008MNRAS.384..107E}; (f) \cite{2008MNRAS.389.1577T}; (g) \cite{2008ApJ...689..377W}; (h) \cite{2009ApJ...704..629M}; (i) \cite{2010ApJ...716..712A}; (j) \cite{2001MNRAS.327.1004B}.
\end{table}

\begin{figure*}
\includegraphics[width=18cm]{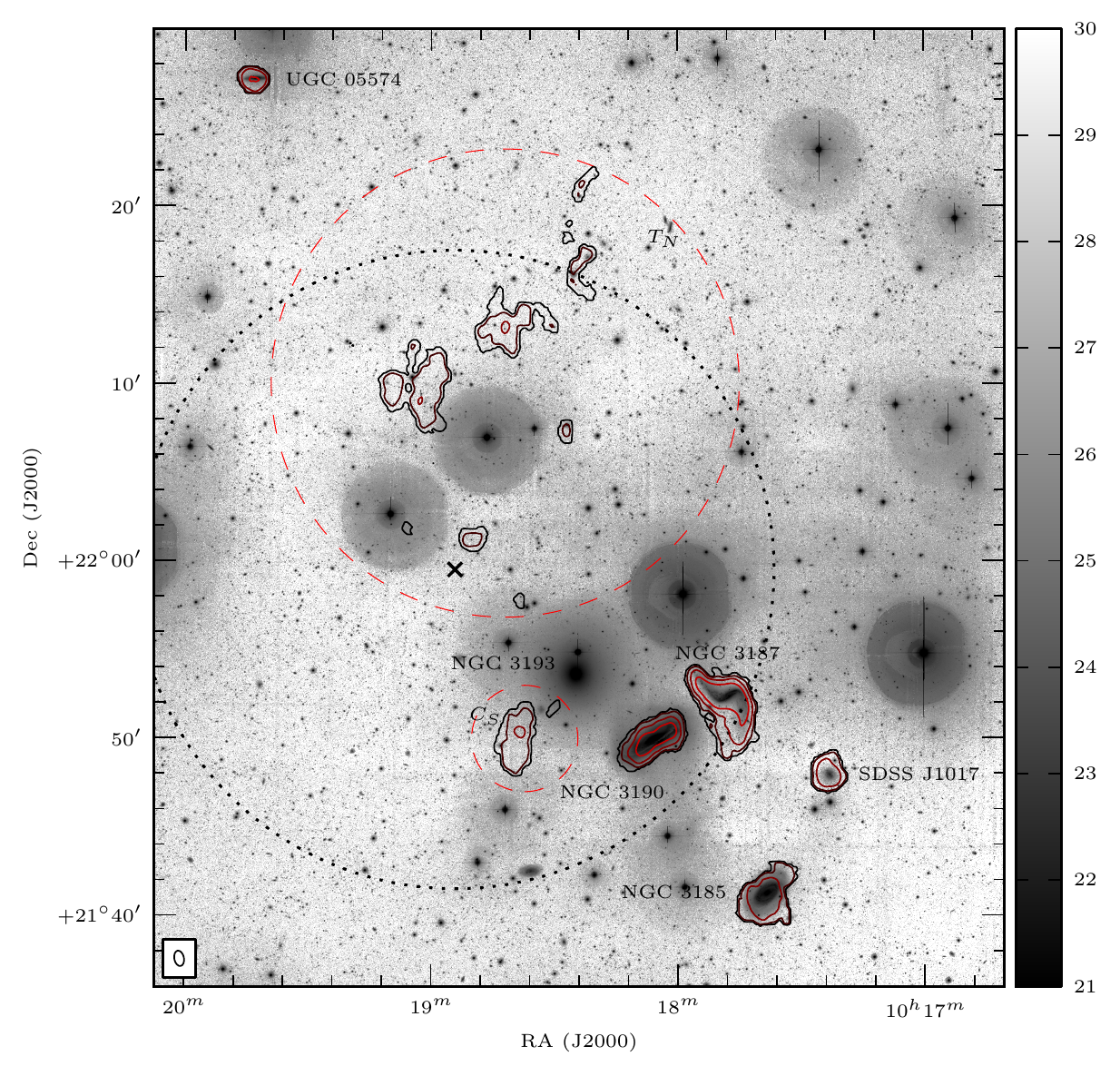}
\caption{Constant-column-density \hi\ contours overlaid on the $g$-band CFHT/MegaCam image. The colourbar on the right is in mag arcsec$^{-2}$. Contours represent $N$(\hi) $=1.0\times10^{19}\times 3^n \ \mathrm{cm}^{-2}$ ($n=0,1,2,3$). Contours are coloured black to red, faint to bright. The black cross indicates the pointing centre of our WSRT observation and the dotted, black circle indicates the primary beam of the WSRT. Large and small dashed, red circles indicate the location of $T_N$ and $C_S$, respectively (see text). The beam of the \hi\ image is shown on the bottom-left corner. Note that UGC~05574 (to the north-east) and SDSS~J1017 (to the south-west) are not members of HCG~44.}
\label{fig:fig01}
\end{figure*}

\begin{figure*}
\includegraphics[width=18cm]{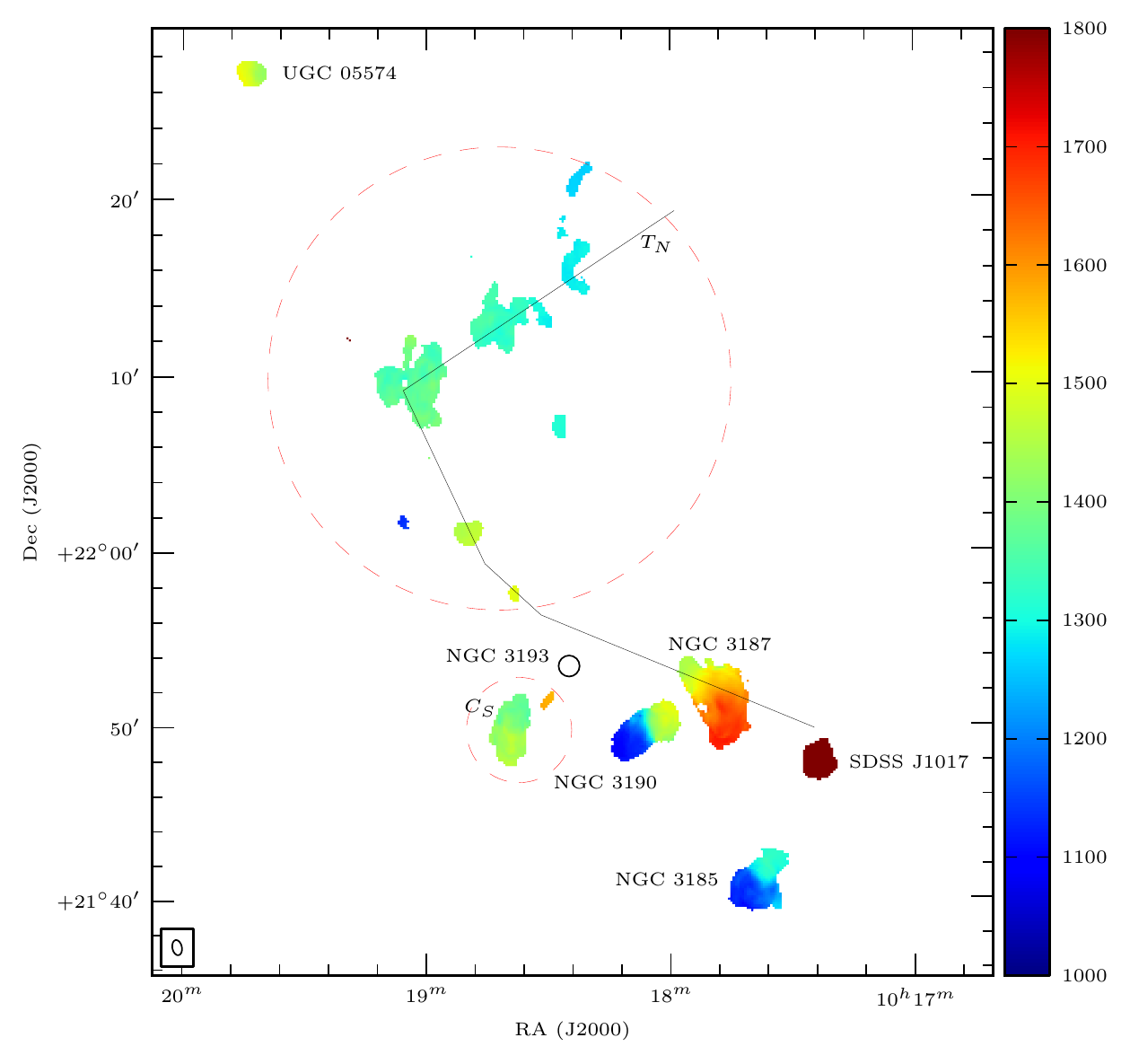}
\caption{Velocity field of the detected \hi. The colourbar on the right is in \kms. Large and small dashed, red circles indicate the location of $T_N$ and $C_S$, respectively. The black open circle indicates the position of NGC~3193, whose recessional velocity from optical spectroscopy is 1381 \kms. The beam of the \hi\ image is shown on the bottom-left corner. The black solid line is the path used to draw the position-velocity diagram shown in Fig. \ref{fig:pv}.}
\label{fig:fig03}
\end{figure*}

Redshift-independent distance estimates of individual galaxies are also puzzling. Taken at face value, the distance of NGC~3193 would put it in the background, as suggested also by \cite{2006A&A...457..771A} based on the non-detection of planetary nebulae in this galaxy. However, error bars on individual distance measurements are large and systematic effects may be important. For example, the two SNIa in NGC~3190 (see table) are extremely obscured and dust corrections are substantial. Furthermore, NGC~3185 has a warp and this would cause a systematic error on the Tully-Fisher distance based on singe-dish \hi\ data.

In this paper we do not attempt to resolve the issue of galaxy distances and group membership. Instead, we assume that all galaxies labelled in Fig. \ref{fig:fig00} belong to the group with the exception of SDSS~J1017, which we consider as a background object based on its large recessional velocity. We assume that the group is at a distance $d_\mathrm{HCG~44}=25$ Mpc. At this distance 15 arcmin correspond to $\sim110$ kpc.

We note that HCG~44 is part of a loose overdensity which includes 10 to 16 bright galaxies depending on the grouping criterion (see group 58 in \citealt{1983ApJS...52...61G} and group 194 in \citealt{1993A&AS..100...47G}). Galaxies in this overdensity have recessional velocity between $\sim1100$ and $\sim1500$ \kms\ and are distributed over a sky area of about $5\times3$ deg$^2$, which corresponds to $2.2\times0.9$ Mpc$^2$ at the distance of HCG~44. Therefore, galaxies in this region are relatively distant from each other (each of them occupies on average 0.6 deg$^2\sim0.1$ Mpc$^2$). In comparison, the galaxy number density within HCG~44 is ten times larger, and this group stands out clearly as a strong and compact overdensity on top of the loose group.

The \hi\ properties of HCG~44 have been studied by \cite{1987ApJS...63..265W} using the Arecibo telescope, \cite{1991AJ....101.1957W} and \cite{2001A&A...377..812V} with Very Large Array (VLA) data, \cite{2001A&A...378..370V} with the Nan\c{c}ay telescope, and \cite{2010ApJ...710..385B} using the Green Bank Telescope (GBT). These studies agree that galaxies in the group (all detected in \hi\ except NGC~3193, which is claimed to be detected by \citealt{2001A&A...378..370V} only) are gas-poor relative to similar objects in the field. \cite{2001A&A...377..812V} estimate the detected \hi\ mass to be just $\sim40$ percent of the expected value for NGC~3187 and $\sim10$ percent for NGC~3185 and NGC~3190 (we revisit these estimates in Sec. \ref{sec:gals}). \cite{2010ApJ...710..385B} report the detection of some of the missing gas as they find excess \hi\ emission in the GBT spectrum\footnote{The GBT beam is $\sim9$ arcmin.} compared to the total VLA spectrum. Based on the strong similarity between the GBT and VLA \hi\ profiles they suggest that the excess gas is dynamically similar to the \hi\ detected by the VLA within individual galaxies in the group.

We observed HCG~44 in \hi\ as part of the multi-wavelength \atlas\ survey \citep{2011MNRAS.413..813C}. The \hi\ observations of this survey were carried out with the Westerbork Synthesis Radio Telescope (WSRT) and presented in \cite{2012MNRAS.422.1835S}. Those observations revealed a few small gas clouds around HCG~44. \cite{2012MNRAS.422.1835S} argue that the distribution of the clouds on the sky and in velocity is suggestive of a long, intra-group \hi\ tail -- the detected clouds being the densest clumps along the hypothetical tail. Here we present new, deeper WSRT observations performed as part of the \atlas\ project to detect the \hi\ tail itself.

\section{Observations}
\label{sec:obs}

\subsection{\hi\ interferometry}
\label{sec:int}

We observed HCG~44 for $6\times12$ h with the WSRT. We pointed the telescope at $\alpha_\mathrm{J2000}= $ 10~h 18~m 54.19~s, $\delta_\mathrm{J2000}= $ 21~d 59~m 30.5~s, which is a position between NGC~3193 and one of the \hi\ clouds detected by \cite{2012MNRAS.422.1835S}. We reduced the data in a standard way using the WSRT pipeline developed by \cite{2012MNRAS.422.1835S}. The \hi\ cube used for our analysis is made using robust=0 weighting and 30 arcsec FWHM tapering. This results in a beam major and minor axis of 53.0 arcsec and 32.7 arcsec, respectively ($\mathrm{PA}=6.5$ deg; the beam axes correspond to 6.4 and 4.0 kpc at the adopted distance, respectively). The cube has velocity resolution of 16 \kms\ after Hanning smoothing. The noise is $\sigma=0.22$ \mJybeam\ in each channel, which corresponds to a formal $5 \sigma$ \hi\ column density sensitivity of $1.1\times10^{19}$ cm$^{-2}$ per resolution element.

We use the source finder described in \cite{2012MNRAS.422.1835S} to detect emission in the \hi\ cube. The finder looks for emission in the original cube and in cubes at different resolutions on the sky and/or in velocity. It flags all voxels above $+n\sigma$ and all voxels below $-n\sigma$ as emission and performs basic size filtering to reduce the noise in the mask at each resolution. Here we use Gaussian filters of FWHM 25 and 50 arcsec on the sky and top-hat filters of width 16, 32, 64, 128, 256 and 384 \kms\ in velocity, and adopt $n=3$. We use the resulting mask to build the total-\hi\ image (which we then correct for the primary beam of the WSRT) and the \hi\ velocity field (intensity-weighted mean) shown in Sec. \ref{sec:res}. All \hi\ flux and mass values reported in this paper are measured from the total-\hi\ image and are therefore corrected for the primary beam.

\subsection{Optical imaging}
\label{sec:megacam}

Deep optical imaging of HCG~44 was obtained with the MegaCam camera mounted on the Canadian-French-Hawaiian Telescope (CFHT). These observations were taken as part of the ATLAS$^\mathrm{3D}$ project \citep{2011MNRAS.413..813C} with the goal of studying early-type galaxies' morphological fine-structure as a relic of their formation \citep{2011MNRAS.417..863D}. We observed the HCG~44 field for $7\times345$ sec in both $g$ and $r$ band applying offsets of $\sim30$ arcmin between consecutive exposures. The seeing was 0.9 and 1.2 arcsec in $g$ and $r$ band, respectively. We refer to \cite{2011MNRAS.417..863D} for a full description of the observations and data reduction. We note here that the observing strategy and data reduction procedures (Elixir-LSB software; Cuillandre et al., in prep.) were chosen to detect stellar structures reaching $\sim28.5$ mag arcsec$^{-2}$ in $g$ band. However, the surface brightness sensitivity may vary from field to field (and within a given field) and we discuss the sensitivity of our image in more detail in the following section.

\section{Results}
\label{sec:res}

\subsection{\hi\ in the intra-group medium}
\label{sec:med}

Figure \ref{fig:fig01} shows \hi\ constant-column-density contours overlaid on the $g$-band CFHT/MegaCam image of HCG~44. The main result of our new observation is the detection of a long \hi\ tail north of NGC~3193 (hereafter $T_N$), which is contained in the large, dashed red circle in the figure. The tail consists of a low-column-density, $\sim20$-arcmin-long component oriented north-west to south-east and a less massive, southern extension. The latter is revealed by two small clouds aligned north-east to south-west in the direction of NGC~3193. We also confirm the \cite{2012MNRAS.422.1835S} detection of a smaller \hi\ complex south-east of NGC~3193 (hereafter $C_S$), indicated by the small, dashed red circle in the figure.

The tail $T_N$ has no diffuse optical counterpart down to the surface brightness sensitivity of the deep image shown in Fig. \ref{fig:fig01} (the same result is obtained inspecting the $r$-band image). In Sec. \ref{sec:megacam} we mentioned that the generic sensitivity of the CFHT/MegaCam images taken as part of the \atlas\ project is $\sim28.5$ mag arcsec$^{-2}$ in $g$ band. In this particular case $T_N$ lays on a relatively clean region of the image, which shows no large-scale noise variations or bright sources with the exception of two stars in the southern part of the tail. The noise level in this region (excluding the two stars) corresponds to a $1.5 \sigma$ detection limit of 28.4 mag arcsec$^{-2}$ within a circular aperture of radius equal to the seeing, consistent with the generic value given above. Such low detection level per resolution element is sufficient considering that we are looking for emission over a scale of many arcminutes.

\begin{figure}
\includegraphics[width=8.4cm]{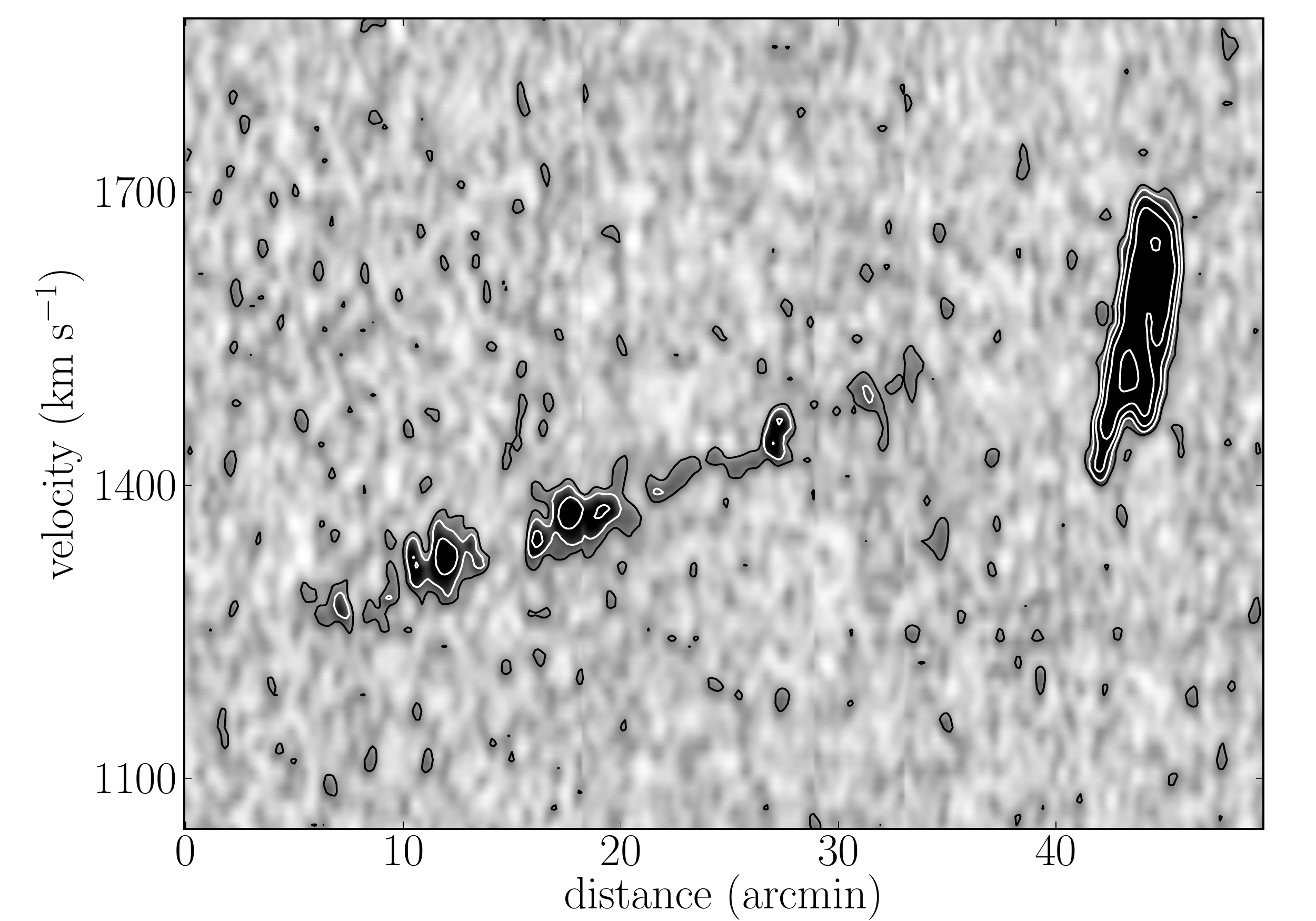}
\caption{Position-velocity diagram of the \hi\ emission along the path shown by the black solid line in Fig. \ref{fig:fig03}. The origin of the horizontal axis is set to be at the northern end of the path. The bright emission at distance $\sim45$ arcmin is \hi\ in NGC~3187.}
\label{fig:pv}
\end{figure}

Figure \ref{fig:fig03} shows the velocity field of the detected \hi. We find a smooth variation of \hi\ recessional velocity along $T_N$. This is shown more clearly by the position-velocity diagram in Fig. \ref{fig:pv}, which is drawn along the path indicated by the black solid line in Fig. \ref{fig:fig03}. The clouds detected by \cite{2012MNRAS.422.1835S} and mentioned in Sec. \ref{sec:intro} are therefore just the tip of the iceberg of a large, coherent, gas distribution, as suggested in that article. Gas in both $T_N$ and $C_S$ is detected at velocities comparable to that of individual galaxies in HCG~44, suggesting that the tails are part of the group. $T_N$ has a total projected length of $\sim220$ kpc at the assumed distance $d_\mathrm{HCG~44}=25$ Mpc .

\begin{figure}
\includegraphics[width=8.4cm]{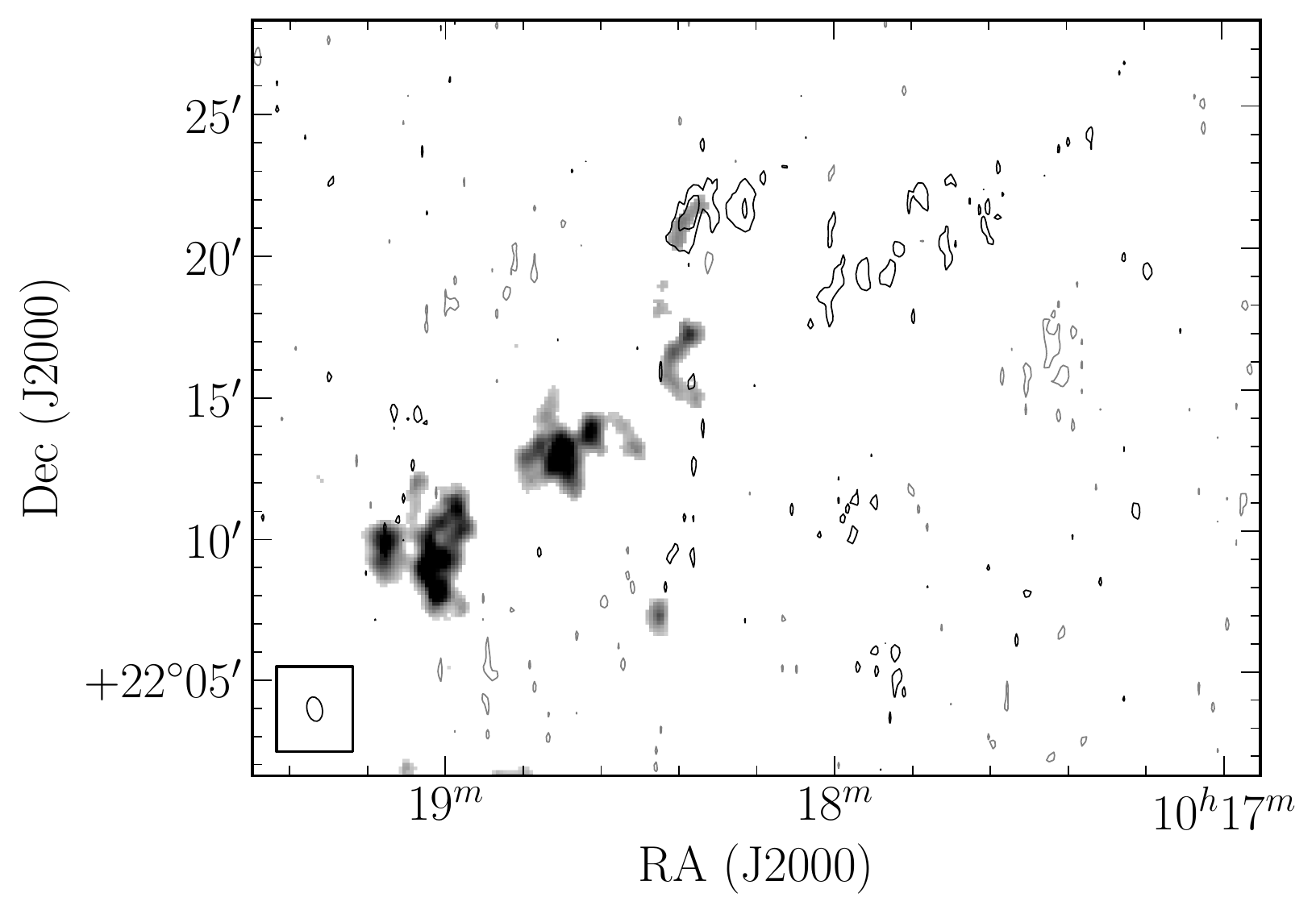}
\caption{Contours of the \hi\ emission in the natural-weighting cube at a recessional velocity of 1248 \kms\ overlaid on a grey-scale total \hi\ image of part of the tail $T_N$. Grey contours are at $-0.4$ \mJybeam, black contours at $+0.4$ and $+0.8$ \mJybeam.  The figure shows possible additional emission north-west of $T_N$. Note that this low-level emission was not cleaned (i.e., it was not deconvolved with the beam).}
\label{fig:tentative}
\end{figure}

Table \ref{tab:hiflux} lists the \hi\ flux of all detected objects shown in Fig. \ref{fig:fig01}. Table \ref{tab:hiflux} also shows the fractional contribution $f$(\hi) of each object to the total \hi\ flux of the group. Assuming that all galaxies in HCG~44 and the intra-group gas are roughly at the same distance we conclude that \hi\ in $T_N$ and $C_S$ amounts to $\sim20$ percent of the total neutral hydrogen mass of of the compact group. This is as much \hi\ as that found in NGC~3190 and almost half the \hi\ mass of NGC~3187. At the assumed distance $d_\mathrm{HCG~44}$ we estimate an \hi\ mass of $4.1\times10^8$ and $1.2\times10^8$ \msun\ for $T_N$ and $C_S$, respectively.

We note that $T_N$ stretches beyond the FWHM of WSRT's primary beam, which is 0.6 deg at the observing frequency of our data. Therefore, it is possible that the gaseous tail is actually longer than shown in Fig. \ref{fig:fig01} and we stop detecting \hi\ because of a decrease in the telescope response. To check whether this is the case we make a natural-weighting \hi\ cube from our WSRT data. This weighting scheme produces a cube with very patchy noise and poorer image quality, but the sensitivity is better than that of the main cube used in this study. Figure \ref{fig:tentative} shows a channel of the natural-weighting cube at velocity 1248 \kms\ suggesting indeed that some additional \hi\ emission may exist north-west of $T_N$.

\begin{figure*}
\includegraphics[width=18cm]{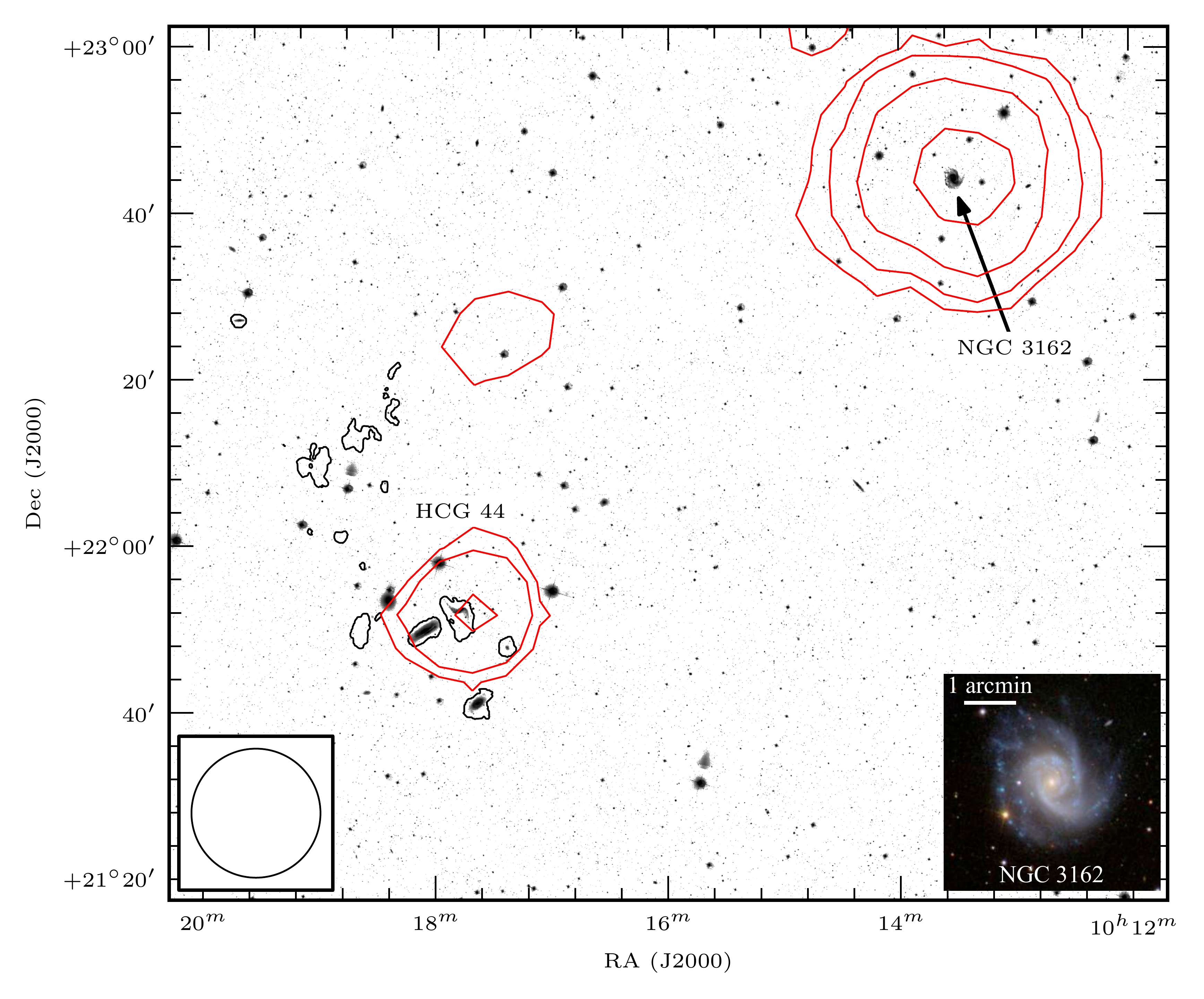}
\caption{HIPASS (red) and WSRT (black) \hi\ contours on top of a SDSS $g$-band image. We show the HIPASS beam (15.5 arcmin) in the bottom-left corner. HIPASS contours are drawn at $N$(\hi) $=1.0\times10^{18}\times 3^n \ \mathrm{cm}^{-2}$ ($n=0,1,2,3$). The WSRT contour is the lowest contour shown in Fig. \ref{fig:fig01}: $1.0\times10^{19}$ cm$^{-2}$. The bottom-right inset shows the SDSS optical colour image of NGC~3162 obtained at http://skyserver.sdss3.org/dr8/en/tools/chart/chart.asp.}
\label{fig:hipass}
\end{figure*}

The existence of a north-western extension of $T_N$ is confirmed by data taken as part of the \hi\ Parkes All Sky Survey (HIPASS, \citealt{2001MNRAS.322..486B}). The HIPASS cube\footnote{available at http://www.atnf.csiro.au/research/multibeam/release .} of this sky area has noise of $\sim15$ \mJybeam, velocity resolution 18 \kms\ and beam FWHM 15.5 arcmin. The formal $5 \sigma$ sensitivity per resolution element is $1.7\times10^{18}$ cm$^{-2}$ for gas filling the beam.  We run the source finder described in Sec. \ref{sec:int} on the cube and show contours of the detected \hi\ in Fig. \ref{fig:hipass} overlaid on a SDSS $g$-band image (we also show the lowest WSRT \hi\ contour; see Fig. \ref{fig:fig01}).

The figure shows two clear HIPASS detections at the location of HCG~44 and NGC~3162 (a spiral galaxy $\sim1.5$ deg north-west of HCG~44). Both detections are listed in the Northern HIPASS catalog by \cite{2006MNRAS.371.1855W} -- objects HIPASS~J1017+21 and HIPASS~J1013+22, respectively. The main part of $T_N$ is visible only at a tentative level in HIPASS (partly because the cube contains many artefacts) and the HIPASS contours in Fig. \ref{fig:hipass} do not show it. On the other hand, the HIPASS cube reveals additional emission (not included in the catalogue of \citealt{2006MNRAS.371.1855W}) just north-west of $T_N$, which was missed by our WSRT data because of its large distance from the pointing centre. The velocity of this emission is consistent with the velocity field shown in Fig. \ref{fig:fig03}. Including this emission, the length of $T_N$ becomes $\sim300$ kpc and its mass $5.2\times10^8$ \msun. This result is confirmed by the newly reduced HIPASS data, which have significantly lower noise level and many less artefacts (Calabretta et al., in prep.). In the new data the full tail is detected clearly (HIPASS team, priv. comm.).

Another interesting aspect revealed by the HIPASS data is that the \hi\ tail stretches towards and possibly connects with NGC~3162 (this galaxy was already detected in \hi\ by, e.g., \citealt{2001A&A...378..370V}). NGC~3162, whose optical image we show in the inset of Fig. \ref{fig:hipass}, has \hi\ recessional velocity of 1300 \kms, comparable to that of HCG~44 members, and is at a projected distance of $\sim650$ kpc from HCG~44 assuming the distance $d_\mathrm{HCG~44}$ along the line of sight. We discuss this finding in Sec. \ref{sec:dis}.

\begin{table}
{\centering
\caption{\hi\ detected with the WSRT in HCG~44}
\begin{tabular}{rrrr}
\hline
object & $F$(\hi) & $f$(\hi) & \mhi \\
  & (Jy km s$^{-1}$) & & ($10^8$ \msun) \\
  (1) & (2) & (3) & (4) \\
 \hline
$T_N$ & 2.79 & 0.15 & 4.1 \\
$C_S$ & 0.82 & 0.05 & 1.2\\
NGC~3185 & 2.11 & 0.12 & 3.1 \\
NGC~3187 & 8.22 & 0.46 & 12.0 \\
NGC~3190 & 4.00 & 0.22 & 5.8 \\
\hline
\multicolumn{4}{c}{\it galaxies not members of HCG~44\rm}\\
SDSS~J1017 & 1.03 & - & 2.0 \\
UGC~05574 & 0.84 & - & 1.1 \\
\hline
\label{tab:hiflux}
\end{tabular}

}

\it Column 1. \rm Object name. \it Column 2. \rm Total \hi\ flux. \it Column 3. \rm Fraction of the total \hi\ flux of HCG~44. \it Column 4. \rm \hi\ mass assuming a distance of $d_\mathrm{HCG~44}=25$ Mpc for galaxies in HCG~44, and Hubble flow distance for galaxies outside HCG~44 (29 and 24 Mpc for SDSS~J1017 and UGC~05574, respectively -- see Table \ref{tab:memb}).
\end{table}

\subsection{\hi\ in galaxies: comparison to previous studies}
\label{sec:gals}

All gas-rich galaxies in Fig. \ref{fig:fig01} were already known to host \hi. \cite{1991AJ....101.1957W} observed this group with the VLA. They report \hi\ fluxes systematically larger than those in Table \ref{tab:hiflux}. Namely, the flux of NGC~3185, NGC~3187, NGC~3190 and SDSS~J1017 is a factor 1.7, 1.3, 1.3 and 2.4 larger than our value, respectively\footnote{These factors include a correction of the VLA fluxes for the primary beam of that telescope, which was not applied by \cite{1991AJ....101.1957W}. The primary-beam correction is of $\sim10$ percent for NGC~3185 and SDSS~J1017 while it is negligible for the other two objects. Note that the VLA flux values are consistent with those reported by \cite{1987ApJS...63..265W} using Arecibo data. They are adopted by \cite{2001A&A...377..812V} in their \hi\ study of Hickson compact groups. They are 5 to 20 percent larger than values in \cite{2001A&A...378..370V}, which have however large uncertainty.}. For NGC~3185 and SDSS~J1017 part of the difference may be due to the better column density sensitivity of the VLA data relative to our WSRT data (by a factor of $\sim1.7$; these galaxies are relatively far from the WSRT pointing centre as shown in Fig. \ref{fig:fig01}). On the contrary, our data are slightly more sensitive at the location of NGC~3187 and NGC~3190 (10 to 20 percent) so at least some of the flux difference must have a different cause.


Flux calibration is not the reason of this discrepancy as we have verified that our calibration is consistent with that of \cite{1991AJ....101.1957W} within $\pm10$ percent. Instead, we do find a relevant difference in the way we build the total-\hi\ image. In our study we use the source finder of \cite{2012MNRAS.422.1835S}, which includes in the \hi\ image both positive and negative noise peaks (see Sec. \ref{sec:int}). On the contrary, \cite{1991AJ....101.1957W} select only voxels above $+1.5 \sigma$. This introduces a positive bias in the total \hi\ flux. Indeed, if we implement their detection criterion in our source finder we obtain fluxes a factor 1.5, 1.1, 1.2, 1.3 larger for the four galaxies mentioned above, respectively. This accounts at least partially for the difference in \hi\ flux. Remaining differences may be explained by the better sensitivity of the VLA data at the location of NGC~3185 and SDSS~J1017 (see above) and by the higher noise level of the VLA data at the location of NGC~3187 and NGC~3190, which would imply a higher level of the bias under discussion.

The \hi\ morphology of galaxies in Fig. \ref{fig:fig01} is in agreement with the image presented by \cite{1991AJ....101.1957W}. The main addition of our deeper image (besides the detection of $T_N$ and $C_S$) is a faint extension of the southern \hi\ warp in NGC~3187. Furthermore, we set a column density limit of $\sim2\times10^{19}$ cm$^{-2}$ on the \hi\ bridge between NGC~3185 and NGC~3190, which was tentatively suggested by \cite{1991AJ....101.1957W} based on the VLA data (this limit takes into account that the two galaxies lay around the half-power point of the WSRT primary beam).

We note that the lower value of our \mhi\ estimates compared to values in \cite{1991AJ....101.1957W} would imply an even larger \hi\ deficiency than that derived by \cite{2001A&A...377..812V} (see Sec. \ref{sec:hcg44}). We revisit the \hi\ deficiency of galaxies in HCG~44 using the relation between \mhi\ and optical diameter $D_{\mathrm{25}, r}$ derived by \cite{2011ApJ...732...93T}. The isophotal major axis of NGC~3185, NGC~3187 and NGC~3190 is given in the SDSS DR7 catalogue and is 8.7, 6.5 and 13.4 kpc, respectively. Given these values, the \hi\ mass predicted by the $1/V_\mathrm{max}$-corrected \mhi-$D_{\mathrm{25}, r}$ relation in \cite{2011ApJ...732...93T} is 7.8, 5.4 and $13.5\times10^9$ \msun, respectively. Therefore, the detected mass of \hi\ is just 4, 22 and 4 percent of the expected value for the three galaxies (6, 3, and $6\ \sigma$ below the expected \hi\ mass, respectively, where $\sigma=0.23$ dex is the r.m.s. residual of \citealt{2011ApJ...732...93T} relation).

\section{Discussion}
\label{sec:dis}

The detection of a 300 kpc-long tail containing $5\times10^8$ \msun\ of \hi\ in a group already observed by various authors with many different radio telescopes may seem surprising (see Sec. \ref{sec:hcg44} for a summary of previous \hi\ observations of HCG~44). Our result demonstrates that signatures of \it on-going \rm galaxy evolution inside groups can be truly elusive. It shows that although \hi\ observations are unique in giving direct evidence of the fundamental role of group processes for galaxy evolution \citep[as shown not only on individual systems but also on large, statistically representative samples; e.g.,][]{2012MNRAS.422.1835S}, such observations need to be very sensitive and cover a large field if we want to gather a complete census of these events. This should be kept in mind when designing future \hi\ surveys. 

Besides the above general conclusion, our observations reveal for the first time direct evidence of gas stripping in HCG~44. This is interesting because galaxies in this group have long been known to be \hi\ deficient and we may be unveiling the cause of at least part of the deficiency. It is therefore interesting to speculate on how the detected \hi\ tail may have formed. How was the gas stripped, and from what galaxy? The goal of the present section is to explore possible answers to these two questions.

\subsection{Formation of the \hi\ tail: ram pressure or tidal stripping}

The two possible mechanisms to form a gas tail within a group are ram pressure and tidal stripping. The former necessitates a dense medium and large relative velocity between the medium and the stripped galaxy. So far, X-ray observations have been unsuccessful in detecting the intra-group medium of HCG~44 (\citealt{2003ApJS..145...39M} using ROSAT data; \citealt{2008MNRAS.388.1245R} using Chandra and XMM-Newton data). \cite{2008MNRAS.388.1245R} estimate that the density of the medium is $n<10^{-4}$ cm$^{-3}$. Analytic calculations indicate that even at such low density some stripping might occur but they also show that this would be a fairly small effect \citep[e.g.,][]{2008MNRAS.388.1245R,2010MNRAS.409.1518F,2011MNRAS.410.2217W}. It therefore seems unlikely that ram pressure is responsible for creating such a prominent tail, longer than and as massive as tails detected in the centre of clusters \citep[e.g.,][]{2005A&A...437L..19O}. Furthermore, no galaxy in the group shows the typical \hi\ morphology caused by ram-pressure stripping -- i.e., gas compressed against the stellar disc on one side of the galaxy and extending to larger radius on the opposite side \citep[e.g.,][]{2004AJ....127.3361K,2007ApJ...659L.115C,2009AJ....138.1741C}. We conclude that ram pressure is an unlikely explanation for the formation of $T_N$.

The alternative hypothesis is that $T_N$ was created by tidal interaction within the group. Tidal forces act on both stars and gas and it may therefore seem surprising that no diffuse stellar light is detected at the location of $T_N$ even in the very deep CFHT/MegaCam image shown in Fig. \ref{fig:fig01}. In fact, observations \citep[e.g.,][]{2004MNRAS.349..922D} and simulations \citep[e.g.,][]{2005MNRAS.357L..21B,2008ApJ...673..787D,2010ApJ...717L.143M} both show that star-less tidal tails can develop following gravitational interaction and depending on the relative distribution of gas and stars in the stripped galaxy. Therefore, tidal interaction is a viable mechanism to create the detected tail and we explore this possibility in the rest of this section.

\subsection{What stripped galaxy?}
\label{sec:whichgal}

The first question we attempt to answer is what galaxy the \hi\ was stripped from. The results described in Sec. \ref{sec:res} suggest two alternative hypotheses. One possibility is that the tail was stripped from NGC~3162 as it flew by the group at high speed (see Fig. \ref{fig:hipass}). This galaxy is currently at a projected distance of $\sim650$ kpc from HCG~44, which it could have covered in $\sim3$ Gyr if it went through the group at a velocity of 200 km/s on the plane of the sky. The optical morphology of this galaxy may support this hypothesis as the stellar disc is lopsided, indicating that it might have been perturbed recently (see inset in Fig. \ref{fig:hipass}). However, lopsidedness is a relatively common phenomenon in spiral galaxies and is no definite proof of an interaction between NGC~3162 and HCG~44. The newly reduced HIPASS data (Calabretta et al., in prep.) hint to a possible \hi\ bridge between $T_N$ and NGC~3162 which, if confirmed, would be a strong clue in favour of this hypothesis. This will be explored further in a future paper presenting the new data.

The other possibility is that the observed \hi\ tail originated from one of the galaxies currently within HCG~44, and NGC~3162 played no significant role. Optical imaging shows that some of the members of HCG~44 have experienced recent tidal interaction (see Fig. \ref{fig:fig00}), and $T_N$ may have formed as part of the same process.

The most obvious candidate for tidal stripping within the group might be NGC~3187. This galaxy exhibits a strong tidal distortion visible as an S-shaped warp and $T_N$ could be seen as an extension of the north-east side of the warp (see Fig. \ref{fig:fig01}; note that any connection between $T_N$ and NGC~3187 would have to be at column density below $\sim10^{19}$ cm$^{-2}$). This is confirmed by the fact that the \hi\ velocity on the south-west end of $T_N$ is similar to that on the north-east side of NGC~3187 (Figs. \ref{fig:fig03} and \ref{fig:pv}). Another interesting point is that gas in $T_N$ amounts to almost half of the total \hi\ mass of NGC~3187. Therefore, tidal stripping would provide a possible explanation for the \hi\ deficiency of this galaxy (see Sec. \ref{sec:gals}). We speculate on the details of the stripping process below.

\subsection{Tidal stripping of NGC~3187}

Numerous previous authors argued that NGC~3187 might be interacting with NGC~3190. While this is possible, it is unlikely that such interaction is responsible for the formation of $T_N$. The reason is that it would be difficult to explain the gap between \hi\ in NGC~3187 and $T_N$. Furthermore, compared to previously known cases of very long tails induced by $\sim$major galaxy interaction and merging \citep[e.g.,][]{1991A&A...243..367M,2011MNRAS.417..863D} the length of the tail -- at least $300$ kpc -- seems too extreme to be caused by this particular galaxy pair.

A mechanism to form very long, low-column-density \hi\ tails (or rings) like that in HCG~44 was proposed by \cite{2005MNRAS.357L..21B}. The main ingredient of their model is the tidal interaction of a low-surface-brightness disc (in this case, NGC~3187) with the gravitational potential of a group of galaxies. This interaction would strip the outer part of the disc (which are often observed to be essentially star-less) and distribute the stripped material on a long tail or ring.

We perform a basic test of this mechanism and estimate the time-scale of the stripping process by building a three-dimensional toy model for the orbit of NGC~3187 around HCG~44 . The model is constrained by the assumed trajectory of NGC~3187 (traced by the \hi\ tail and the current position of the galaxy) and its velocity relative to the group. In this model we assume the group potential to be fixed and Keplerian. We take the centre of the potential to be at the position of NGC~3193 and place it at coordinates $[x,y,z]=[0,0,0]$ (we discuss the orientation of the sky relative to these three axes below). We further assume that NGC~3187 follows a parabolic orbit parallel to the $xy$ plane (and therefore at constant $z=z_0$).

For $z_0=0$ the orbit is determined only by the total mass of the system $M_\mathrm{tot}$ and the focal length of the parabola $f$. However, in this model we assume that the orbital plane is displaced relative to the centre of the group ($z_0\neq 0$; we show below that this is necessary for the parabola to fit the data). Under this assumption NGC~3187 should move also along the $z$ axis but we neglect this effect. Such additional motion would be little constrained by the available data and would not in any case have a significant impact on the time-scale of the orbit, which is our primary goal. Therefore, the orbit remains perfectly planar (this approximation becomes increasingly wrong as the $xy$ distance of NGC~3187 from the group centre gets closer to the value of $z_0$).

\begin{figure}
\includegraphics[width=8.4cm]{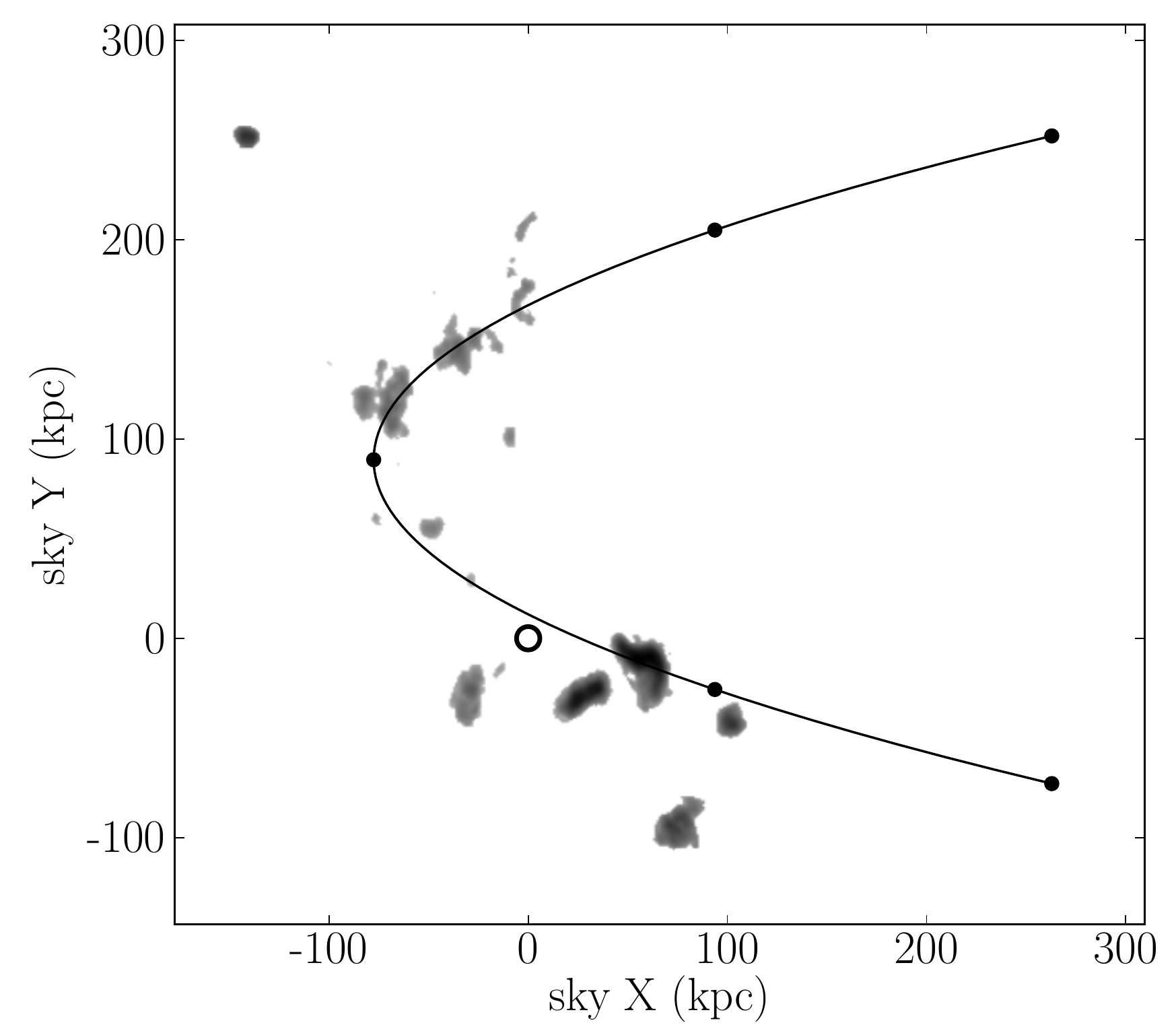}
\caption{Total \hi\ image of HCG~44 (logarithmic greyscale) with superimposed the possible orbit of NGC~3187 through the group (black line; see text). The open black circle indicates the position of NGC~3193, assumed centre of the group. Solid black circles are spaced at 0.5 Gyr time intervals.}
\label{fig:fig05}
\end{figure}

The toy model described above has only three free parameters: $M_\mathrm{tot}$, $f$ and $z_0$. We fix $M_\mathrm{tot}=5\times10^{12}$ \msun. We vary $f$ between 50 and 200 kpc, and $z_0$ between 0 and 200 kpc in an attempt to reproduce the observations. Furthermore, in order to compare the model to the data we need to project it on the sky. This introduces additional degrees of freedom. We make use of 3D visualisation to find a favourable projection.

Figure \ref{fig:fig05} shows a possible orbit of NGC~3187, which reproduces the location of the \hi\ tail and that of the galaxy at once. The orbit is obtained with $f=75$ kpc and $z_0=100$ kpc, and is viewed at an inclination of 60 deg away from face-on (note that this is not the result of a formal fit and is therefore just one possible solution of the problem). An important feature of the model is that the predicted velocity of NGC~3187 relative to the group centre (i.e., NGC~3193) is 235 \kms\ along the line of sight. This is a good match to the observed value of 210 \kms\ (see Table \ref{tab:memb}). Therefore, this simple model can explain the anomalous, high velocity of NGC~3187 relative to other members of HCG~44.

Within the context of this simple model NGC~3187 passed the vertex of the orbit $\sim400$ Myr ago and was at the current location of the north-west end of the \hi\ tail (revealed by the HIPASS data shown in Fig. \ref{fig:hipass}) $\sim900$ Myr ago. This suggests that the galaxy may have been stripped of the \hi\ now in $T_N$ ($\sim1/3$ of its initial \hi\ mass) within a time interval shorter than 1 Gyr. It is interesting that this timescale is consistent with the $\sim2$ Gyr timescale for quenching of star formation in galaxy groups recently derived by \cite{2012arXiv1208.1762R}.

There are some important differences between HCG~44 and the systems simulated by \cite{2005MNRAS.357L..21B}. Firstly, their simulations  predict the existence of a leading tail which, in the case of NGC~3187, should start at the southern tip of the warp, bend towards north-west and eventually join $T_N$ to form an intra-group \hi\ ring. This is not observed. Bekki et al.'s is, however, an idealised model. In reality this process occurs in the presence of many other galaxies. It is possible that such second tail was destroyed by an interaction with NGC~3190 (which is tidally disturbed) and gas was instead dispersed towards the location of $C_S$. This idea is supported by both the \hi\ and optical morphology of NGC~3187's southern tidal tail. Figure \ref{fig:fig01} shows that this tail broadens and bends towards NGC~3190, giving support to the idea that the two galaxies are interacting. Therefore, it is possible that also gas in $C_S$ once belonged to NGC~3187, and the above estimate on the amount of gas stripped from this galaxy should be regarded as a lower limit.

Another difference is that in Bekki et al.'s model the stripped galaxy is not tidally distorted, unlike NGC~3187. Again, interaction with NGC~3190 may explain the difference (note that, as we have argued above, this interaction is unlikely to be the formation mechanism of $T_N$ itself).

We stress that we view this model as a very simple (but nevertheless useful) description of a possible formation mechanism for the detected tail. However, the system is complex: the group membership is still to be established and a number of galaxies might be playing a relevant role, including the very distant NGC~3162 (see Fig. \ref{fig:hipass} and Sec. \ref{sec:whichgal}). As a consequence, substantial differences between the data and idealised simulations like that of \cite{2005MNRAS.357L..21B} or simple models like the parabolic trajectory presented above are not a surprise. More detailed modelling is beyond the scope of this paper (for example, a realistic mass model of the group would be needed, and the line-of-sight velocity of gas in $T_N$ should be used as a constraint for the model -- taking into account also rotation within the stripped galaxy). Such modelling or the exploration of different formation mechanisms for the tail (for example interaction between NGC~3162 and a member of HCG~44 within the group potential, as in the models by \citealt{2005MNRAS.363L..21B}) would strongly benefit from having deep \hi\ observations over a larger field.

\section{Conclusions}
\label{sec:con}

We have presented deep \hi\ and optical imaging of the galaxy group HCG~44 obtained with the WSRT and CFHT/MegaCam, respectively, as part of the \atlas\ project. We detect a long intra-group \hi\ tail and additional intra-group gas which, together, amount to 20 percent of the \hi\ mass of the group. Combining these data with archive HIPASS observations we find that the main tail contains $\sim5\times10^8$ \msun\ of \hi\ and is $\sim300$ kpc long assuming a distance of 25 Mpc. The tail has no diffuse optical counterpart down to $\sim28.5$ mag arcsec$^{-2}$ in $g$ band.

We discuss viable formation mechanisms for the \hi\ tail. Based on the available data (including X-ray imaging) it is unlikely that the tail is caused by ram pressure stripping. Instead, it is possible that the \hi\ was stripped from NGC~3187, a member of HCG~44, by the group tidal field. We present a simple model for the orbit of the stripped galaxy through the group. Within this model, NGC~3187 has been stripped of $1/3$ of its initial gas mass in less than 1 Gyr. This is consistent with recent estimates of the timescale for quenching of star formation in galaxy groups ($\sim2$ Gyr).

The proposed model is intentionally simple and it is possible that other processes have contributed to shaping the properties of galaxies in the group (e.g., tidal interaction between group members and, possibly, some ram-pressure stripping). Another possibility suggested by HIPASS data is that \hi\ in the tail was stripped from NGC~3162, a spiral galaxy now 650 kpc from the group. Future work will investigate this possibility using \hi\ data of better quality than those available at the moment.

Regardless of the precise formation mechanism, the detected \hi\ tail is the first, direct evidence of gas stripping in HCG~44, a group long known to be deficient of \hi. Our result highlights the importance of group processing as a driver of galaxy evolution, but also the observational challenge that has to be overcome in order to detect signatures of these processes. Sensitive \hi\ observations over a large field are needed to gather a complete census of this kind of events in the local Universe.

\section*{Acknowledgments}

PS acknowledges useful discussions with T. van Albada, J. van Gorkom, R. Sancisi, S. Tonnesen and C. Toribio.
  
MC acknowledges support from a Royal Society University Research Fellowship. 
This work was supported by the rolling grants `Astrophysics at Oxford' PP/E001114/1 and ST/H002456/1 and visitors grants PPA/V/S/2002/00553, PP/E001564/1 and ST/H504862/1 from the UK Research Councils. RLD acknowledges travel and computer grants from Christ Church, Oxford and support from the Royal Society in the form of a Wolfson Merit Award 502011.K502/jd. RLD also acknowledges the support of the ESO Visitor Programme which funded a 3 month stay in 2010. 
SK acknowledges support from the the Royal Society Joint Projects Grant JP0869822. 
RMcD is supported by the Gemini Observatory, which is operated by the Association of Universities for Research in Astronomy, Inc., on behalf of the international Gemini partnership of Argentina, Australia, Brazil, Canada, Chile, the United Kingdom, and the United States of America. 
LMD acknowledges support from the Lyon Institute of Origins under grant ANR-10-LABX-66. 
TN and MBois acknowledge support from the DFG Cluster of Excellence `Origin and Structure of the Universe'. 
MS acknowledges support from a STFC Advanced Fellowship ST/F009186/1. 
PS is a NWO/Veni fellow. 
(TAD) The research leading to these results has received funding from the European Community's Seventh Framework Programme (/FP7/2007-2013/) under grant agreement No 229517. 
MBois has received, during this research, funding from the European Research Council under the Advanced Grant Program Num 267399-Momentum. 
The authors acknowledge financial support from ESO. 
This research has made use of the NASA/IPAC Extragalactic Database (NED) which is operated by the Jet Propulsion Laboratory, California Institute of Technology, under contract with the National Aeronautics and Space Administration. 
This paper is based on observations obtained with the Westerbork Synthesis Radio Telescope, which is operated by the ASTRON (Netherlands Foundation for Research in Astronomy) with support from the Netherlands Foundation for Scientific Research NWO, and with MegaPrime/MegaCam, a joint project of CFHT and CEA/DAPNIA, at the CFHT, which is operated by the National Research Council (NRC) of Canada, the Institute National des Sciences de lÕUnivers of the Centre National de la Recherche Scientifique of France and the University of Hawaii.

\bibliographystyle{mn2e}
\bibliography{hcg44}

\begin{thebibliography}{68}
\expandafter\ifx\csname natexlab\endcsname\relax\def\natexlab#1{#1}\fi

\bibitem[{{Aguerri} {et~al}\mbox{.}(2006){Aguerri}, {Castro-Rodr{\'{\i}}guez},
  {Napolitano}, {Arnaboldi}, \& {Gerhard}}]{2006A&A...457..771A}
{Aguerri} J.~A.~L., {Castro-Rodr{\'{\i}}guez} N., {Napolitano} N., {Arnaboldi}
  M., {Gerhard} O., 2006, \aap, 457, 771

\bibitem[{{Amanullah} {et~al}\mbox{.}(2010){Amanullah}, {Lidman}, {Rubin},
  {Aldering}, {Astier}, {Barbary}, {Burns}, {Conley}, {Dawson}, {Deustua},
  {Doi}, {Fabbro}, {Faccioli}, {Fakhouri}, {Folatelli}, {Fruchter}, {Furusawa},
  {Garavini}, {Goldhaber}, {Goobar}, {Groom}, {Hook}, {Howell}, {Kashikawa},
  {Kim}, {Knop}, {Kowalski}, {Linder}, {Meyers}, {Morokuma}, {Nobili},
  {Nordin}, {Nugent}, {{\"O}stman}, {Pain}, {Panagia}, {Perlmutter}, {Raux},
  {Ruiz-Lapuente}, {Spadafora}, {Strovink}, {Suzuki}, {Wang}, {Wood-Vasey},
  {Yasuda}, \& {Supernova Cosmology Project}}]{2010ApJ...716..712A}
{Amanullah} R. {et~al.}, 2010, \apj, 716, 712

\bibitem[{{Barnes} {et~al}\mbox{.}(2001){Barnes}, {Staveley-Smith}, {de Blok},
  {Oosterloo}, {Stewart}, {Wright}, {Banks}, {Bhathal}, {Boyce}, {Calabretta},
  {Disney}, {Drinkwater}, {Ekers}, {Freeman}, {Gibson}, {Green}, {Haynes}, {te
  Lintel Hekkert}, {Henning}, {Jerjen}, {Juraszek}, {Kesteven}, {Kilborn},
  {Knezek}, {Koribalski}, {Kraan-Korteweg}, {Malin}, {Marquarding}, {Minchin},
  {Mould}, {Price}, {Putman}, {Ryder}, {Sadler}, {Schr{\"o}der}, {Stootman},
  {Webster}, {Wilson}, \& {Ye}}]{2001MNRAS.322..486B}
{Barnes} D.~G. {et~al.}, 2001, \mnras, 322, 486

\bibitem[{{Bekki} \& {Couch}(2011)}]{2011MNRAS.415.1783B}
{Bekki} K., {Couch} W.~J., 2011, \mnras, 415, 1783

\bibitem[{{Bekki}, {Koribalski} \& {Kilborn}(2005){Bekki}, {Koribalski}, \&
  {Kilborn}}]{2005MNRAS.363L..21B}
{Bekki} K., {Koribalski} B.~S., {Kilborn} V.~A., 2005, \mnras, 363, L21

\bibitem[{{Bekki} {et~al}\mbox{.}(2005){Bekki}, {Koribalski}, {Ryder}, \&
  {Couch}}]{2005MNRAS.357L..21B}
{Bekki} K., {Koribalski} B.~S., {Ryder} S.~D., {Couch} W.~J., 2005, \mnras,
  357, L21

\bibitem[{{Blakeslee} {et~al}\mbox{.}(2001){Blakeslee}, {Lucey}, {Barris},
  {Hudson}, \& {Tonry}}]{2001MNRAS.327.1004B}
{Blakeslee} J.~P., {Lucey} J.~R., {Barris} B.~J., {Hudson} M.~J., {Tonry}
  J.~L., 2001, \mnras, 327, 1004

\bibitem[{{Borthakur}, {Yun} \& {Verdes-Montenegro}(2010){Borthakur}, {Yun}, \&
  {Verdes-Montenegro}}]{2010ApJ...710..385B}
{Borthakur} S., {Yun} M.~S., {Verdes-Montenegro} L., 2010, \apj, 710, 385

\bibitem[{{Cappellari} {et~al}\mbox{.}(2011{\natexlab{a}}){Cappellari},
  {Emsellem}, {Krajnovi{\'c}}, {McDermid}, {Scott}, {Verdoes Kleijn}, {Young},
  {Alatalo}, {Bacon}, {Blitz}, {Bois}, {Bournaud}, {Bureau}, {Davies}, {Davis},
  {de Zeeuw}, {Duc}, {Khochfar}, {Kuntschner}, {Lablanche}, {Morganti}, {Naab},
  {Oosterloo}, {Sarzi}, {Serra}, \& {Weijmans}}]{2011MNRAS.413..813C}
{Cappellari} M. {et~al.}, 2011{\natexlab{a}}, \mnras, 413, 813

\bibitem[{{Cappellari} {et~al}\mbox{.}(2011{\natexlab{b}}){Cappellari},
  {Emsellem}, {Krajnovi{\'c}}, {McDermid}, {Serra}, {Alatalo}, {Blitz}, {Bois},
  {Bournaud}, {Bureau}, {Davies}, {Davis}, {de Zeeuw}, {Khochfar},
  {Kuntschner}, {Lablanche}, {Morganti}, {Naab}, {Oosterloo}, {Sarzi}, {Scott},
  {Weijmans}, \& {Young}}]{2011MNRAS.416.1680C}
{Cappellari} M. {et~al.}, 2011{\natexlab{b}}, \mnras, 416, 1680

\bibitem[{{Chung} {et~al}\mbox{.}(2009){Chung}, {van Gorkom}, {Kenney},
  {Crowl}, \& {Vollmer}}]{2009AJ....138.1741C}
{Chung} A., {van Gorkom} J.~H., {Kenney} J.~D.~P., {Crowl} H., {Vollmer} B.,
  2009, \aj, 138, 1741

\bibitem[{{Chung} {et~al}\mbox{.}(2007){Chung}, {van Gorkom}, {Kenney}, \&
  {Vollmer}}]{2007ApJ...659L.115C}
{Chung} A., {van Gorkom} J.~H., {Kenney} J.~D.~P., {Vollmer} B., 2007, \apjl,
  659, L115

\bibitem[{{Davies} {et~al}\mbox{.}(2004){Davies}, {Minchin}, {Sabatini}, {van
  Driel}, {Baes}, {Boyce}, {de Blok}, {Disney}, {Evans}, {Kilborn}, {Lang},
  {Linder}, {Roberts}, \& {Smith}}]{2004MNRAS.349..922D}
{Davies} J. {et~al.}, 2004, \mnras, 349, 922

\bibitem[{{Dressler}(1980)}]{1980ApJ...236..351D}
{Dressler} A., 1980, \apj, 236, 351

\bibitem[{{Duc} \& {Bournaud}(2008)}]{2008ApJ...673..787D}
{Duc} P.-A., {Bournaud} F., 2008, \apj, 673, 787

\bibitem[{{Duc} {et~al}\mbox{.}(2011){Duc}, {Cuillandre}, {Serra},
  {Michel-Dansac}, {Ferriere}, {Alatalo}, {Blitz}, {Bois}, {Bournaud},
  {Bureau}, {Cappellari}, {Davies}, {Davis}, {de Zeeuw}, {Emsellem},
  {Khochfar}, {Krajnovi{\'c}}, {Kuntschner}, {Lablanche}, {McDermid},
  {Morganti}, {Naab}, {Oosterloo}, {Sarzi}, {Scott}, {Weijmans}, \&
  {Young}}]{2011MNRAS.417..863D}
{Duc} P.-A. {et~al.}, 2011, \mnras, 417, 863

\bibitem[{{Elias-Rosa} {et~al}\mbox{.}(2008){Elias-Rosa}, {Benetti}, {Turatto},
  {Cappellaro}, {Valenti}, {Arkharov}, {Beckman}, {di Paola}, {Dolci},
  {Filippenko}, {Foley}, {Krisciunas}, {Larionov}, {Li}, {Meikle},
  {Pastorello}, {Valentini}, \& {Hillebrandt}}]{2008MNRAS.384..107E}
{Elias-Rosa} N. {et~al.}, 2008, \mnras, 384, 107

\bibitem[{{English} {et~al}\mbox{.}(2010){English}, {Koribalski},
  {Bland-Hawthorn}, {Freeman}, \& {McCain}}]{2010AJ....139..102E}
{English} J., {Koribalski} B., {Bland-Hawthorn} J., {Freeman} K.~C., {McCain}
  C.~F., 2010, \aj, 139, 102

\bibitem[{{Freeland}, {Sengupta} \& {Croston}(2010){Freeland}, {Sengupta}, \&
  {Croston}}]{2010MNRAS.409.1518F}
{Freeland} E., {Sengupta} C., {Croston} J.~H., 2010, \mnras, 409, 1518

\bibitem[{{Freeland}, {Stilp} \& {Wilcots}(2009){Freeland}, {Stilp}, \&
  {Wilcots}}]{2009AJ....138..295F}
{Freeland} E., {Stilp} A., {Wilcots} E., 2009, \aj, 138, 295

\bibitem[{{Garcia}(1993)}]{1993A&AS..100...47G}
{Garcia} A.~M., 1993, A\&AS, 100, 47

\bibitem[{{Geller} \& {Huchra}(1983)}]{1983ApJS...52...61G}
{Geller} M.~J., {Huchra} J.~P., 1983, \apjs, 52, 61

\bibitem[{{Hibbard} {et~al}\mbox{.}(2001){Hibbard}, {van Gorkom}, {Rupen}, \&
  {Schiminovich}}]{2001ASPC..240..657H}
{Hibbard} J.~E., {van Gorkom} J.~H., {Rupen} M.~P., {Schiminovich} D., 2001, in
  Astronomical Society of the Pacific Conference Series, Vol. 240, Gas and
  Galaxy Evolution, {Hibbard} J.~E., {Rupen} M., {van Gorkom} J.~H., eds., p.
  657

\bibitem[{{Hickson}(1982)}]{1982ApJ...255..382H}
{Hickson} P., 1982, \apj, 255, 382

\bibitem[{{Kenney}, {van Gorkom} \& {Vollmer}(2004){Kenney}, {van Gorkom}, \&
  {Vollmer}}]{2004AJ....127.3361K}
{Kenney} J.~D.~P., {van Gorkom} J.~H., {Vollmer} B., 2004, \aj, 127, 3361

\bibitem[{{Kern} {et~al}\mbox{.}(2008){Kern}, {Kilborn}, {Forbes}, \&
  {Koribalski}}]{2008MNRAS.384..305K}
{Kern} K.~M., {Kilborn} V.~A., {Forbes} D.~A., {Koribalski} B., 2008, \mnras,
  384, 305

\bibitem[{{Kilborn} {et~al}\mbox{.}(2009){Kilborn}, {Forbes}, {Barnes},
  {Koribalski}, {Brough}, \& {Kern}}]{2009MNRAS.400.1962K}
{Kilborn} V.~A., {Forbes} D.~A., {Barnes} D.~G., {Koribalski} B.~S., {Brough}
  S., {Kern} K., 2009, \mnras, 400, 1962

\bibitem[{{Koribalski}, {Gordon} \& {Jones}(2003){Koribalski}, {Gordon}, \&
  {Jones}}]{2003MNRAS.339.1203K}
{Koribalski} B., {Gordon} S., {Jones} K., 2003, \mnras, 339, 1203

\bibitem[{{Koribalski} \& {L{\'o}pez-S{\'a}nchez}(2009)}]{2009MNRAS.400.1749K}
{Koribalski} B.~S., {L{\'o}pez-S{\'a}nchez} {\'A}.~R., 2009, \mnras, 400, 1749

\bibitem[{{Kreckel} {et~al}\mbox{.}(2012){Kreckel}, {Platen},
  {Arag{\'o}n-Calvo}, {van Gorkom}, {van de Weygaert}, {van der Hulst}, \&
  {Beygu}}]{2012AJ....144...16K}
{Kreckel} K., {Platen} E., {Arag{\'o}n-Calvo} M.~A., {van Gorkom} J.~H., {van
  de Weygaert} R., {van der Hulst} J.~M., {Beygu} B., 2012, \aj, 144, 16

\bibitem[{{Krumm} \& {Burstein}(1984)}]{1984AJ.....89.1319K}
{Krumm} N., {Burstein} D., 1984, \aj, 89, 1319

\bibitem[{{Mandel} {et~al}\mbox{.}(2009){Mandel}, {Wood-Vasey}, {Friedman}, \&
  {Kirshner}}]{2009ApJ...704..629M}
{Mandel} K.~S., {Wood-Vasey} W.~M., {Friedman} A.~S., {Kirshner} R.~P., 2009,
  \apj, 704, 629

\bibitem[{{Meurer} {et~al}\mbox{.}(1996){Meurer}, {Carignan}, {Beaulieu}, \&
  {Freeman}}]{1996AJ....111.1551M}
{Meurer} G.~R., {Carignan} C., {Beaulieu} S.~F., {Freeman} K.~C., 1996, \aj,
  111, 1551

\bibitem[{{Michel-Dansac} {et~al}\mbox{.}(2010){Michel-Dansac}, {Duc},
  {Bournaud}, {Cuillandre}, {Emsellem}, {Oosterloo}, {Morganti}, {Serra}, \&
  {Ibata}}]{2010ApJ...717L.143M}
{Michel-Dansac} L. {et~al.}, 2010, \apjl, 717, L143

\bibitem[{{Mirabel}, {Lutz} \& {Maza}(1991){Mirabel}, {Lutz}, \&
  {Maza}}]{1991A&A...243..367M}
{Mirabel} I.~F., {Lutz} D., {Maza} J., 1991, \aap, 243, 367

\bibitem[{{Morganti} {et~al}\mbox{.}(2006){Morganti}, {de Zeeuw}, {Oosterloo},
  {McDermid}, {Krajnovi{\'c}}, {Cappellari}, {Kenn}, {Weijmans}, \&
  {Sarzi}}]{2006MNRAS.371..157M}
{Morganti} R. {et~al.}, 2006, \mnras, 371, 157

\bibitem[{{Mulchaey} {et~al}\mbox{.}(2003){Mulchaey}, {Davis}, {Mushotzky}, \&
  {Burstein}}]{2003ApJS..145...39M}
{Mulchaey} J.~S., {Davis} D.~S., {Mushotzky} R.~F., {Burstein} D., 2003, \apjs,
  145, 39

\bibitem[{{Oosterloo} \& {van Gorkom}(2005)}]{2005A&A...437L..19O}
{Oosterloo} T., {van Gorkom} J., 2005, \aap, 437, L19

\bibitem[{{Oosterloo} {et~al}\mbox{.}(2007){Oosterloo}, {Morganti}, {Sadler},
  {van der Hulst}, \& {Serra}}]{2007A&A...465..787O}
{Oosterloo} T.~A., {Morganti} R., {Sadler} E.~M., {van der Hulst} T., {Serra}
  P., 2007, \aap, 465, 787

\bibitem[{{Postman} \& {Geller}(1984)}]{1984ApJ...281...95P}
{Postman} M., {Geller} M.~J., 1984, \apj, 281, 95

\bibitem[{{Rasmussen} {et~al}\mbox{.}(2012{\natexlab{a}}){Rasmussen}, {Bai},
  {Mulchaey}, {van Gorkom}, {Jeltema}, {Zabludoff}, {Wilcots}, {Martini},
  {Lee}, \& {Roberts}}]{2012ApJ...747...31R}
{Rasmussen} J. {et~al.}, 2012{\natexlab{a}}, \apj, 747, 31

\bibitem[{{Rasmussen} {et~al}\mbox{.}(2012{\natexlab{b}}){Rasmussen},
  {Mulchaey}, {Bai}, {Ponman}, {Raychaudhury}, \&
  {Dariush}}]{2012arXiv1208.1762R}
{Rasmussen} J., {Mulchaey} J.~S., {Bai} L., {Ponman} T.~J., {Raychaudhury} S.,
  {Dariush} A., 2012{\natexlab{b}}, ArXiv e-prints

\bibitem[{{Rasmussen} {et~al}\mbox{.}(2008){Rasmussen}, {Ponman},
  {Verdes-Montenegro}, {Yun}, \& {Borthakur}}]{2008MNRAS.388.1245R}
{Rasmussen} J., {Ponman} T.~J., {Verdes-Montenegro} L., {Yun} M.~S.,
  {Borthakur} S., 2008, \mnras, 388, 1245

\bibitem[{{Reindl} {et~al}\mbox{.}(2005){Reindl}, {Tammann}, {Sandage}, \&
  {Saha}}]{2005ApJ...624..532R}
{Reindl} B., {Tammann} G.~A., {Sandage} A., {Saha} A., 2005, \apj, 624, 532

\bibitem[{{Sancisi} {et~al}\mbox{.}(2008){Sancisi}, {Fraternali}, {Oosterloo},
  \& {van der Hulst}}]{2008A&ARv..15..189S}
{Sancisi} R., {Fraternali} F., {Oosterloo} T., {van der Hulst} T., 2008, A\&AR,
  15, 189

\bibitem[{{Schiminovich} {et~al}\mbox{.}(1994){Schiminovich}, {van Gorkom},
  {van der Hulst}, \& {Kasow}}]{1994ApJ...423L.101S}
{Schiminovich} D., {van Gorkom} J.~H., {van der Hulst} J.~M., {Kasow} S., 1994,
  \apjl, 423, L101

\bibitem[{{Schneider} {et~al}\mbox{.}(1983){Schneider}, {Helou}, {Salpeter}, \&
  {Terzian}}]{1983ApJ...273L...1S}
{Schneider} S.~E., {Helou} G., {Salpeter} E.~E., {Terzian} Y., 1983, \apjl,
  273, L1

\bibitem[{{Scott} {et~al}\mbox{.}(2012){Scott}, {Cortese}, {Brinks},
  {Bravo-Alfaro}, {Auld}, \& {Minchin}}]{2012MNRAS.419L..19S}
{Scott} T.~C., {Cortese} L., {Brinks} E., {Bravo-Alfaro} H., {Auld} R.,
  {Minchin} R., 2012, \mnras, 419, L19

\bibitem[{{Serra} {et~al}\mbox{.}(2012){Serra}, {Oosterloo}, {Morganti},
  {Alatalo}, {Blitz}, {Bois}, {Bournaud}, {Bureau}, {Cappellari}, {Crocker},
  {Davies}, {Davis}, {de Zeeuw}, {Duc}, {Emsellem}, {Khochfar},
  {Krajnovi{\'c}}, {Kuntschner}, {Lablanche}, {McDermid}, {Naab}, {Sarzi},
  {Scott}, {Trager}, {Weijmans}, \& {Young}}]{2012MNRAS.422.1835S}
{Serra} P. {et~al.}, 2012, \mnras, 422, 1835

\bibitem[{{Springob} {et~al}\mbox{.}(2009){Springob}, {Masters}, {Haynes},
  {Giovanelli}, \& {Marinoni}}]{2009ApJS..182..474S}
{Springob} C.~M., {Masters} K.~L., {Haynes} M.~P., {Giovanelli} R., {Marinoni}
  C., 2009, \apjs, 182, 474

\bibitem[{{Szomoru} {et~al}\mbox{.}(1996){Szomoru}, {van Gorkom}, {Gregg}, \&
  {Strauss}}]{1996AJ....111.2150S}
{Szomoru} A., {van Gorkom} J.~H., {Gregg} M.~D., {Strauss} M.~A., 1996, \aj,
  111, 2150

\bibitem[{{Takanashi}, {Doi} \& {Yasuda}(2008){Takanashi}, {Doi}, \&
  {Yasuda}}]{2008MNRAS.389.1577T}
{Takanashi} N., {Doi} M., {Yasuda} N., 2008, \mnras, 389, 1577

\bibitem[{{Thilker} {et~al}\mbox{.}(2009){Thilker}, {Donovan}, {Schiminovich},
  {Bianchi}, {Boissier}, {Gil de Paz}, {Madore}, {Martin}, \&
  {Seibert}}]{2009Natur.457..990T}
{Thilker} D.~A. {et~al.}, 2009, \nat, 457, 990

\bibitem[{{Toribio} {et~al}\mbox{.}(2011){Toribio}, {Solanes}, {Giovanelli},
  {Haynes}, \& {Martin}}]{2011ApJ...732...93T}
{Toribio} M.~C., {Solanes} J.~M., {Giovanelli} R., {Haynes} M.~P., {Martin}
  A.~M., 2011, \apj, 732, 93

\bibitem[{{Tully}(1988)}]{1988ngc..book.....T}
{Tully} R.~B., 1988, {Nearby galaxies catalog}

\bibitem[{{van der Hulst}(1979)}]{1979A&A....75...97V}
{van der Hulst} J.~M., 1979, \aap, 75, 97

\bibitem[{{van Driel} {et~al}\mbox{.}(2001){van Driel}, {Marcum}, {Gallagher},
  {Wilcots}, {Guidoux}, \& {Monnier Ragaigne}}]{2001A&A...378..370V}
{van Driel} W., {Marcum} P., {Gallagher}, III J.~S., {Wilcots} E., {Guidoux}
  C., {Monnier Ragaigne} D., 2001, \aap, 378, 370

\bibitem[{{van Driel} \& {van Woerden}(1991)}]{1991A&A...243...71V}
{van Driel} W., {van Woerden} H., 1991, \aap, 243, 71

\bibitem[{{Verdes-Montenegro} {et~al}\mbox{.}(2001){Verdes-Montenegro}, {Yun},
  {Williams}, {Huchtmeier}, {Del Olmo}, \& {Perea}}]{2001A&A...377..812V}
{Verdes-Montenegro} L., {Yun} M.~S., {Williams} B.~A., {Huchtmeier} W.~K., {Del
  Olmo} A., {Perea} J., 2001, \aap, 377, 812

\bibitem[{{Verheijen} \& {Zwaan}(2001)}]{2001ASPC..240..867V}
{Verheijen} M.~A.~W., {Zwaan} M., 2001, in Astronomical Society of the Pacific
  Conference Series, Vol. 240, Gas and Galaxy Evolution, {Hibbard} J.~E.,
  {Rupen} M., {van Gorkom} J.~H., eds., p. 867

\bibitem[{{Vorontsov-Velyaminov}(1959)}]{1959VV....C......0V}
{Vorontsov-Velyaminov} B.~A., 1959, in Atlas and catalog of interacting
  galaxies (1959), p.~0

\bibitem[{{Wang} {et~al}\mbox{.}(2006){Wang}, {Wang}, {Pain}, {Zhou}, \&
  {Li}}]{2006ApJ...645..488W}
{Wang} X., {Wang} L., {Pain} R., {Zhou} X., {Li} Z., 2006, \apj, 645, 488

\bibitem[{{Westmeier}, {Braun} \& {Koribalski}(2011){Westmeier}, {Braun}, \&
  {Koribalski}}]{2011MNRAS.410.2217W}
{Westmeier} T., {Braun} R., {Koribalski} B.~S., 2011, \mnras, 410, 2217

\bibitem[{{Williams}, {McMahon} \& {van Gorkom}(1991){Williams}, {McMahon}, \&
  {van Gorkom}}]{1991AJ....101.1957W}
{Williams} B.~A., {McMahon} P.~M., {van Gorkom} J.~H., 1991, \aj, 101, 1957

\bibitem[{{Williams} \& {Rood}(1987)}]{1987ApJS...63..265W}
{Williams} B.~A., {Rood} H.~J., 1987, \apjs, 63, 265

\bibitem[{{Wilman} {et~al}\mbox{.}(2009){Wilman}, {Oemler}, {Mulchaey},
  {McGee}, {Balogh}, \& {Bower}}]{2009ApJ...692..298W}
{Wilman} D.~J., {Oemler}, Jr. A., {Mulchaey} J.~S., {McGee} S.~L., {Balogh}
  M.~L., {Bower} R.~G., 2009, \apj, 692, 298

\bibitem[{{Wong} {et~al}\mbox{.}(2006){Wong}, {Ryan-Weber}, {Garcia-Appadoo},
  {Webster}, {Staveley-Smith}, {Zwaan}, {Meyer}, {Barnes}, {Kilborn},
  {Bhathal}, {de Blok}, {Disney}, {Doyle}, {Drinkwater}, {Ekers}, {Freeman},
  {Gibson}, {Gurovich}, {Harnett}, {Henning}, {Jerjen}, {Kesteven}, {Knezek},
  {Koribalski}, {Mader}, {Marquarding}, {Minchin}, {O'Brien}, {Putman},
  {Ryder}, {Sadler}, {Stevens}, {Stewart}, {Stootman}, \&
  {Waugh}}]{2006MNRAS.371.1855W}
{Wong} O.~I. {et~al.}, 2006, \mnras, 371, 1855

\bibitem[{{Wood-Vasey} {et~al}\mbox{.}(2008){Wood-Vasey}, {Friedman}, {Bloom},
  {Hicken}, {Modjaz}, {Kirshner}, {Starr}, {Blake}, {Falco}, {Szentgyorgyi},
  {Challis}, {Blondin}, {Mandel}, \& {Rest}}]{2008ApJ...689..377W}
{Wood-Vasey} W.~M. {et~al.}, 2008, \apj, 689, 377

\end{thebibliography}

\label{lastpage}

\end{document}